\documentclass{aa}

\usepackage{graphicx}
\usepackage{txfonts}
\usepackage{lipsum}
\usepackage{natbib}
\usepackage{booktabs}
\usepackage{subcaption}         
\usepackage{lscape}             
\usepackage{placeins}           

\usepackage{keyval}
\usepackage{hyperref}

\begin{document}

   \title{Stellar encounters in the solar neighbourhood and the special case of GJ~710}


   \author{Eloi Fernandez-Puig\inst{1,2}
        \and Juan Carlos Morales\inst{1,2}
        \and Ignasi Ribas \inst{1,2}
        \and J\'ulia Laguna-Miralles\inst{1,3}
        \and Pol Guijosa\inst{1,4}}

   \institute{
            Institut de Ciències de l’Espai (ICE, CSIC), Campus UAB, C/ Can Magrans s/n, 08193 Bellaterra, Spain
            \and Institut d'Estudis Espacials de Catalunya (IEEC), C/Esteve Terradas, 1, Edifici RDIT, Campus PMT-UPC, E-08860 Castelldefels (Barcelona), Spain
            \and Institute of Astronomy, University of Cambridge, Madingley Rd, Cambridge CB3 0HA, UK
            \and Dept. Ingeniería Energética, Universidad Nacional de Educación a Distancia (UNED), C/ Juan del Rosal 12, 28040 Madrid, Spain}

   \date{Accepted, April 2026}

  \abstract{}
    {We present a comprehensive characterisation of close stellar encounters in the solar vicinity, with a particular focus on placing the predicted fly-by of GJ~710 in context. This star will come extremely close ($0.0621$\,pc or $\sim10^4$\,AU) to the Solar System in approximately 1.3\,Myr.}
    {Using a linear motion approximation, we identified past and future close stellar encounters within 1\,pc of the Solar System, using a complete sample of nearby stars. We assessed the completeness of our dataset and applied corrections to the radial velocities, accounting for gravitational redshift and convective blueshift. Such effects can bias the measured velocities and affect the derived encounter parameters. Furthermore, we computed close encounters for all the stars in the Solar System vicinity to build a statistically significant sample of such events. We accounted for binary systems and common proper motion pairs, applying corrections to account for incompleteness at the edges of our time window.}
    {We derived reliable statistics for close stellar encounters of stars within 25\,pc of the Sun. We report a rate of encounters within 1\,pc and within 0.47\,Myr of $10.6 \pm 4.5$ per Myr and star, implying an average of one encounter every $95^{+71}_{-28}$\,kyr. Additionally, from the resulting distribution, we can evaluate the relative frequency of the upcoming GJ~710 fly-by, which is found to be quite rare, at a rate of one per $\sim50$\,Myr.}
    {This work provides new insights into the dynamic environment of the solar neighbourhood by quantifying the rates and distributions of close stellar encounter events.}

   \maketitle
   \nolinenumbers

\section{Introduction}

Planetary systems do not exist in isolation. They reside within a galaxy, where gravitational interactions with neighbouring stars, as well as catastrophic events such as supernovae and gamma-ray bursts, can influence their long-term stability and habitability conditions \citep{Pyne2024}. Estimating the rate and properties of stellar encounters has significant implications for both stellar dynamics and planetary system evolution. Stellar close encounters with the Sun have been studied by numerous authors \citep{Rickman1976, Matthews1994, Weissman1996, Dehnen1998, Garcia-Snchez2001, Levison2004, Jimnez-Torres2011, Bailer-Jones2015, Higuchi2015, Feng&BailerJones2015, Mamajek2015, Berski2016}, using data from the \texttt{Hipparcos} mission \citep{Perryman1997}. However, these studies have been limited by the completeness of the catalogues employed, leading to an estimated rate of $11.7\pm1.3$ stellar encounters per Myr within 1\,pc of the Sun \citep{Garcia-Snchez2001}.

The first data release from \textit{Gaia}  \citep[DR1;][]{GaiaCollaboration2016a, GaiaCollaboration2016b} revolutionised this field as the mission was able to detect and measure nearly all of the local stellar systems within 50\,pc of the Sun. Several authors recalculated close encounters based on the new data \citep{Berski2016, Bobylev2017, Torres2018, Bailer-Jones2018}. These works were superseded by the following release of the \textit{Gaia} catalogue DR2 \citep{GaiaCollaboration2018}, which provided an additional 7.2 million radial velocity (RV) measurements, improving the accuracy of the parameters of close encounters \citep{Bailer-Jonesetal2018, Bobylev2020}. With the \textit{Gaia} DR2 data, \citet{Bailer-Jonesetal2018} estimated the rate of encounters within 1\,pc (and within 15\,Myr) to be $19.7\pm2.2$\,Myr$^{-1}$, after applying incompleteness corrections.

The most recent \textit{Gaia} release, DR3 \citep{Vallenari2023}, provides RVs for 34 million bright stars. Using the new data, \cite{Bailer-Jones2022} identified several close encounters (within 1\,pc of the Sun) for the first time, and the encounter times, distances, and velocities of previously known close encounters were determined more precisely on account of the significantly improved precision of \textit{Gaia} DR3 over earlier releases. One such Solar System close encounter is the remarkable case of GJ~710. Several works have reported on it in the past, noting that the star was predicted pass by at only $\sim$0.06\,pc (just over 2 light-months) of the Solar System in approximately 1.3 million years \citep{Berski2016, Feng2019, Bailer-Jones2022}. The star is currently located at a distance of 19.09\,pc, but the small value of the proper motion combined with the negative RV indicates that it is moving directly towards the Solar System.

A close stellar passage could induce gravitational perturbations on the outer regions of the Solar System, and potentially even influence planetary orbits or cause cometary ``showers'' into the inner planets \citep{Rickman1976, Weissman1996}. Several studies \citep{Malmberg2011, LiDaohai2019, LiDaohaiII2020} have addressed how close stellar fly-bys affect the long-term stability of the Solar System or drive episodes of increased bombardment on Earth. Nevertheless, a clear link between fly-bys and the latter is yet to emerge \citep{Zeebe2025}. \citet{Kaib2025} suggest that stellar encounters substantially increase the risk of planetary instability compared to isolated system models. These instabilities are more likely to result in systems losing multiple planets.

The impact of close stellar encounters on the orbits of Oort cloud bodies has also been considered \citep[see e.g.][]{Oort1950, Biermann1983, Weissman1996}. \citeauthor{Torres2019} investigated such influences and concluded that although individual encounters rarely alter cometary orbits, the cumulative effect of encounters within 1\,pc can strongly perturb the Oort Cloud. In a compact Oort-cloud configuration, where comet semi-major axes are confined within $0.25$ pc, the cumulative interactions of encounters dominate over the effect of the Galactic tide, leading to the ejection of comets out to interstellar space. The passage of GJ~710, in particular, will produce a major perturbation to the inner Oort Cloud. These results indicate that external stellar perturbations play a relevant role in the long-term erosion of Oort Cloud–like structures, both in the Solar System and in other systems. Thus, under the hypothesis that other planetary systems also possess structures similar to the Oort cloud, this leads us to assume a large population of cometary bodies in the interstellar space, potentially explaining the origin of interstellar visitors such as 1I/Oumuamua, 2I/Borisov, and 3I/ATLAS \citep{Torres2019, Guo2025, PerezCouto2025}.

Previous works on close stellar encounters have mainly focused on the Sun, whereas the nature and distribution of such events for other nearby stars remain largely unexplored. Considering a larger sample of stars would enable a more precise statistical analysis of such events and, at the same time, to place the Sun in the context of the statistical distribution. With the excellent quality \textit{Gaia} astrometric data at hand, the main limitation that persists is related to the absence of precise RV data for many stars. Although galactic models are often used to estimate correction factors, complementary spectroscopic RV surveys can significantly improve the precision of encounter parameters. Understanding the rate and characteristics of such stellar interactions is essential for developing a complete picture of stellar dynamics in the solar neighbourhood and to comprehend planetary system evolution, as well as the persistence of long-term habitable conditions.

In this work we re-evaluate the past and future close stellar encounters of the Sun using \textit{Gaia} astrometry data combined with RVs from ground-based surveys and we perform a statistical analysis of stellar encounters in the solar vicinity. Our goal is to to contextualise the Sun’s encounter history relative to nearby stars and to assess the possible frequency of fly-bys that come as close as the prediction for GJ~710 in the Sun's local environment. In Sect.\,\ref{sec:Sample}, we describe the sample of stars used for our statistical studies. Section\,\ref{sec:methodology} describes our methodology to estimate stellar close encounters and Sect.\,\ref{sec:results} presents our results. Finally, in Sect.\,\ref{sec:conclusions}, we summarise the conclusions of the present work.

\section{Sample} \label{sec:Sample}
For this study, we made use of the most recent DR3 \textit{Gaia} data release \citep{Vallenari2023, Prusti2016, Babusiaux2022}. We limited our sample to stars located within 100 parsecs of the Sun with a parallax-error ratio greater than 5, corresponding to a parallax uncertainty of less than 20\%. For each star, we extracted the right ascension (RA), declination (Dec), parallax, proper motions in right ascension (pmRA) and declination (pmDec), RVs, and their associated uncertainties. Additionally, we extracted the effective temperature (teff\_gspphot), surface gravity (logg\_gspphot), radius (radius\_gspphot), and the spectral type (sectraltype\_esphs) for the gravitational redshift and the convective blueshift corrections.

To complement \textit{Gaia} astrometry with additional high-accuracy RVs, we cross-matched the sample with the \texttt{carmencita} catalogue \citep{Caballero2016}, which contains RV measurements from the CARMENES instrument \citep{Lafarga2020} and other sources for 2,180 M dwarfs. Since M dwarfs constitute nearly 80\% of all stars in the Galaxy, improving the reliability of their RVs data represents a substantial enhancement. We preferably selected, when available, RVs from \citet{Lafarga2020}, followed by those listed in the \texttt{carmencita} catalogue (when unavailable, we used \textit{Gaia} DR3 instead). This cross-matching added 212 new RVs and allowed us to correct 1,502 existing values from the \textit{Gaia} DR3 catalogue, significantly increasing the precision of the final dataset.

To calculate the absolute RV, two additional effects of opposite signs need to be considered: the convective blueshift and the gravitational redshift. Previous studies (e.g. \citealt{Bailer-Jones2022}) have argued that applying such corrections to the RVs is not necessary because they are typically smaller than the \textit{Gaia} DR3 RV uncertainties and depend on stellar parameters that are typically not well constrained. However, we reckon that these corrections should be included. For example, the median gravitational redshift for M dwarfs (the most common stellar type) is $\sim0.6$\,km s$^{-1}$, comparable to most \textit{Gaia} RV uncertainties and twice the median RV uncertainty from \texttt{carmencita}. Conversely, convective blueshift for M dwarfs does not exceed $-0.1$\,km s$^{-1}$ \citep{Liebing2021}, which does not compensate fully the gravitational redshift and leads to a net positive RV. These effects introduce systematic biases that could be particularly significant for close-encounter computations involving the Sun. Therefore, we preferred to apply the corrections in spite of their considerable complexity.

We applied a gravitational redshift correction to the RV to all stars in our sample. For objects with available stellar radius and surface gravity ($log\ g$) measurements, we computed the correction directly. For stars lacking these parameters but with a known spectral type, we adopted as a proxy, the median correction value corresponding to stars of the same spectral type. The $log\  g$ values from \texttt{carmencita} were obtained by computing them from the mass and radius values when available. For stars whose spectral type was labelled as 'unknown' or not provided at all, we assigned the global median correction derived from all stars with a computed correction. This approach is justified by the fact that our dataset is distance-limited (within 100 pc) and most targets are expected to be main sequence stars instead of giants, for which this assumption would not hold.
 
Additionally, we applied a convective blueshift correction following the empirical relation presented in \cite{Liebing2021}. This correction was applied only to stars classified as F, G, K, or M types. In accordance with the fit shown in their Fig. 6, we set the correction to zero for stars with effective temperature below 4100 K or above 6000 K. As with the gravitational redshift correction, for stars lacking an effective temperature measurement, we adopted the median correction corresponding to their spectral type as a proxy value.

We pay particular attention here to GJ~710, whose physical and dynamical properties are listed in Table\,\ref{tab:Gl710_basicdata}. We computed the absolute RV from the CARMENES spectra by using {\tt RACCOON} \cite{Lafarga2020}, which measures the RV by cross-correlating each spectrum with a pre-defined mask. The resulting RV is $-14.4493 \pm 0.0004$\,km\,s$^{-1}$. Gravitational redshift and convective blueshift corrections were calculated as explained above, using the effective temperature, surface gravity and radius values provided in the \texttt{carmencita} catalogue. We obtained a gravitational redshift correction of $0.645\pm0.015$\,km\,s$^{-1}$ and a convective blueshift of $-0.100\pm0.016$\,km\,s$^{-1}$, resulting in an absolute RV of $-13.905 \pm 0.022$\,km s$^{-1}$.

\begin{table}
\centering
\caption{Basic physical and dynamical properties of GJ~710.}
\label{tab:Gl710_basicdata}
\begin{tabular}{lll}
\hline
\textbf{Parameter} & \textbf{Value} & \textbf{Refs.} \\
\hline
Alternative ID           & HD 168442                    &   \\
CARMENES ID              & Karmn J18198-019             &   \\ 
RA, Dec (J2000)          & 18:19:50.84, $-$01:56:19.00  & 2 \\
$\varpi$ (mas)           & $52.396 \pm 0.0171$          & 2 \\
pmRA (mas yr$^{-1}$)     & $-0.414 \pm 0.019$           & 2 \\
pmDE (mas yr$^{-1}$)     & $-0.108 \pm 0.017$           & 2 \\
$V_r$ (km s$^{-1}$)      & $-14.4493 \pm 0.0004$        & 1 \\
${V_r}^{\rm abs}$ (km s$^{-1}$) & $-13.905 \pm 0.022$   & 1 \\
Spectral type            & K7\,V                        & 2 \\
$T_{\mathrm{eff}}$ (K)   & $4155 \pm 14$                & 1 \\
Mass (M$_{\odot}$)       & $0.595 \pm 0.013$            & 1 \\
Radius (R$_{\odot}$)     & $0.5865 \pm 0.0043$          & 1 \\
Luminosity (L$_{\odot}$) & $0.09236 \pm 0.00051$        & 1 \\
\hline
\end{tabular}
\vspace{2mm}
\begin{minipage}{0.9\textwidth}
\small
\vspace{2mm}
\textbf{Refs.} (1) This work; (2) \cite{Vallenari2023}
\end{minipage}
\end{table}

The apparent brightness decrease of stars with distance make them less likely to be detected or measured with precision, resulting in a degradation of the catalogue completeness. This should not be a major concern for \textit{Gaia} data, but our sample is further limited by the availability of measured RVs. Since our aim is to carry out a statistical study of close encounters between stars, we evaluated the completeness of our sample by obtaining the density number of stars as a function of distance, computed from the \textit{Gaia} $G$-band magnitude and the distance.

We found that the stellar density number significantly decreases after $\sim$20\,pc, which suggests that this is the full completeness limit. By extrapolating the 20-pc stellar number density, we estimate that at $50$\,pc ,our catalogue contains $91\%$ of the stars and at 100\,pc the completeness ratio decreases to $69\%$. With the goal of finding compromise between completeness and having a dataset sufficiently numerous to yield statistically significant results, we restricted our study of encounters among stars (other than the Sun) to stars within 50\,pc of the Sun. This subsample consists of a total of 28\,107 stars with available RV measurements.

\section{Close encounter analysis}\label{sec:methodology}

The close encounter calculations were performed by assuming the linear motion approximation (LMA), which assumes that stars move at constant velocities along straight-line trajectories. This approach does not consider the acceleration from the Galactic potential, but it was selected because the primary objective of our work is to conduct a broad statistical assessment of encounter rates and characteristics, which can then be compared to the Sun's case; rather than determining high-precision periastron parameters. The simplification avoids expensive dynamical calculations. Furthermore, we show below that the effect of considering the Galactic potential in our calculations is negligible given the time and distance intervals explored.

Under the assumption of uniform motion, a star's position as a function of time is given by \(\vec{r}(t) = \vec{r_0} + \vec{v}t\), where $\vec{r_0}$ is the initial position at $t=0$. The time of closest distance or periastron time, $t_{\mathrm{p}}$, obtained by minimising $|\vec{r}(t)|^2$), is given by
\begin{equation}\label{perih_time}
    t_{\mathrm{p}} = -\frac{\vec{r_0}\cdot\vec{v}}{v^2}.
\end{equation}
\noindent Substituting $t_{\mathrm{p}}$ in the expression for $\vec{r}(t)$ results in
\begin{equation}\label{perih_distance}
    d_{\mathrm{p}} = \left(\vec{r_0} - t_{\mathrm{p}}^2 v^2\right)^{1/2},
\end{equation}
\noindent which is the minimum distance, or periastron distance between the two stellar systems. For the close encounter calculations involving the Sun, it is more convenient to work in equatorial coordinates. In this case, the periastron distance can be expressed in terms of radial and transversal velocities as
\begin{equation}\label{Sun_perih_distance}
    d_{\mathrm{p}} = |\vec{r_0}| \sqrt{1-\frac{v_{r}^2}{v^2}} = |\vec{r_0}| \frac{v_{\mathrm{tr}}}{|v|},
\end{equation}
\noindent where $|\vec{r_0}| = 1/\varpi$ is the inverse of the parallax angle ($\varpi$), $v_r$ is the RV with respect to the Sun and $v_{\mathrm{tr}}$ is the transversal velocity obtained from the star's proper motion.

For close encounters among stars (other than the Sun), a transformation from equatorial celestial coordinates to the Cartesian Galactic frame was performed, using RA, Dec, parallax, proper motions, and RVs of the stars to compute the position and velocity as $\vec{r_0} = (x, y, z)$ and $\vec{v} = (u, v, w)$ \citep{Johnson1987}. The periastron speed ($v_{\mathrm{p}}$) is defined as the relative velocity at the moment of closest approach. Since stellar velocities remain constant under the LMA, $v_{\mathrm{p}}$ is computed as the modulus of the relative velocity vector, $|\vec{v}|$. In the case of close encounters with the Sun, the periastron speed is (as expected) very similar to the radial component (RV) of the velocity vector.

In our analysis we define a close encounter as a stellar passage occurring within 1\,pc distance. For reference, the average spacing between stellar systems in the solar neighbourhood is approximately 2.2\,pc, and the Sun’s current nearest neighbour, the $\alpha$ Centauri system, lies just 1.3\,pc away. For further perspective, the Solar System's Oort cloud extends out to 0.25--0.50\,pc from the Sun \citep{Brasser2012}. As demonstrated by \citeauthor{Torres2019}, the cumulative effect of passing stars within 1\,pc can perturb the comets in the Oort cloud, significantly influencing their dynamics. This potential for cometary disturbance is the main reason why 1\,pc was chosen as the threshold for a close encounter in our work.

To ensure consistency and avoid statistical biases when comparing Solar System close encounters with those among other stellar systems, we adopted uniform volume limits across analyses. For Solar System encounters, we considered two cases: (i) stars within 100\,pc to facilitate comparison with previous studies and (ii) stars within 25\,pc where we can ensure a high completeness level of the catalogue. Encounters among other stars were computed considering spheres of radius $l_{\mathrm{lim}} = 25$\,pc centred on targets located in the inner sphere of 25\,pc around the Sun. By construction, this effectively considers stars up to 50\,pc from the Sun, where the catalogue completeness remains high. This matched 25\,pc scheme ensures statistically comparable volumes, thereby avoiding statistical biases. This approach also substantially reduces computational demands.

The spatial boundary naturally introduces a time limit, as a star requires a finite time to cross the $l_{\mathrm{lim}}$ radius sphere at a constant velocity. Consequently, the characteristic timescale, $t_{\mathrm{lim}}$, was defined as the time a star would take to travel a distance $l_{\mathrm{lim}}$ with a velocity corresponding to the 90th percentile of the RV distribution, which results in values of $t_{\mathrm{lim}}=1.85$\,Myr and 0.47\,Myr for 100\,pc and 25\,pc boundaries, respectively. This introduces a time window for close encounter detection ranging from $-t_{\mathrm{lim}}$ to $t_{\mathrm{lim}}$.  

To evaluate the number of close encounters with the Solar System, taking into account the star coordinates, proper motions, and RV uncertainty, we first computed the periastron distance and the corresponding time of closest approach using Eqs. \eqref{perih_time} and \eqref{perih_distance}. For computational efficiency, we ran an initial encounter calculations and adopted an initial upper limit of 1.1\,pc on the periastron distance. For stars with a periastron distance below this threshold, we performed an uncertainty analysis using a Monte Carlo error propagation method. To identify all potential close encounters comprehensively, we classified events as close encounters if the lower  1-$\sigma$ confidence bound on periastron distance fell below 1\,pc. In these cases, we proceeded to compute additional encounter parameters, such as the periastron speed. However, cases where the 1$\sigma$ uncertainty exceeds 50\% of the nominal periastron distance were deemed unreliable and thus excluded. This selective process ensured that computational resources were focused only on reliable, high-confidence encounters.

We also evaluated the validity of the LMA, which rests on the negligible effect of the Galactic gravitational potential over the limited spatial and temporal scales of our analysis. To quantify this, we considered the axisymmetric potential of \cite{Miyamoto1975}. The maximum differential acceleration experienced by a star within 50\,pc of the Sun is $\sim 4\times 10^{-2}$\,pc\,Myr$^{-2}$, representing only a 0.6\% difference of the Galactic acceleration at the solar position. When propagated over the relevant timescales ($t_{\mathrm{lim}}$), this acceleration induces a position shift of approximately $10^{-2}$\,pc in the Galactic radial component. Since this is two orders of magnitude smaller than our 1\,pc encounter threshold and does not systematically bias our statistics, we concluded that incorporating a full Galactic potential was unnecessary.

For the GJ~710--Sun encounter, the Galactic gravitational potential cannot be neglected, as the encounter reaches a minimum distance comparable to $10^{-2}$\,pc and the time integration is longer. To accurately determine this minimum distance, we performed a dedicated numerical integration of the trajectories of both the Sun and GJ~710 using the Runge-Kutta (RK4) method and considering accelerations derived from the  Miyamoto-Nagai potential (see Table~\ref{tab:MN_potential}) for the parameter values). This is expressed as

\begin{align}\label{eq:MN-potential}
\Phi_d &= -\frac{GM}{\sqrt{R^2 + \left(a + \sqrt{z^2 + b^2}\right)^2}}, \nonumber\\
\Phi_{b/h} &= -\frac{GM}{\sqrt{R^2 + z^2 + b^2}} \nonumber,\\
\Phi &= \Phi_d + \Phi_{b} + \Phi_h,
\end{align}

\begin{table}[t]
\centering
\caption{Parameters of Miyamoto-Nagai axisymmetric potential components for the Milky Way.}
\label{tab:MN_potential}
\begin{tabular}{lcccc}
\hline
\textbf{Component} & \textbf{M ($10^{10}$ M$_\odot$)} & \textbf{a (pc)} & \textbf{b (pc)} \\
\hline
Disk     & 7.91     & 3500  & 250   \\
Bulge    & 1.4     & -      & 350    \\
Halo     & 69.8    & -      & 24000   \\
\hline
\end{tabular}
\end{table}

\section{Results and discussion}
\label{sec:results}

\subsection{Close encounters with the Sun}

\begin{figure}
        \includegraphics[width=\columnwidth]{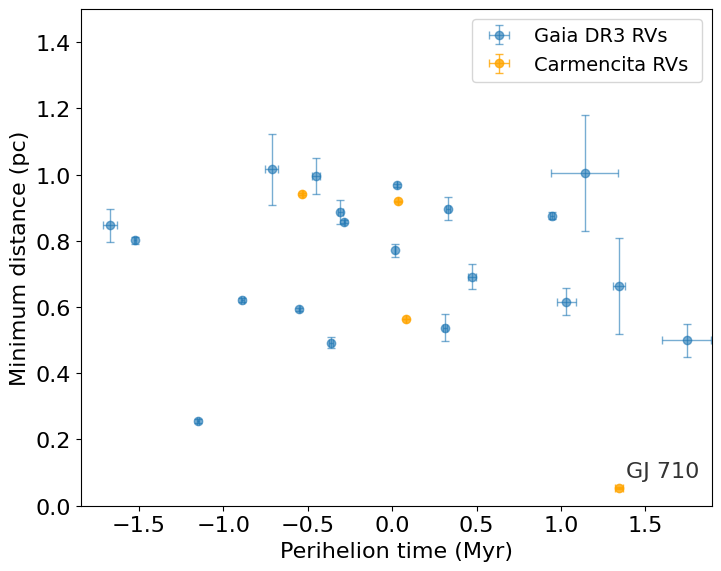}
    \caption{Close encounters of the Sun with stars within 100\,pc using RVs from \textit{Gaia} DR3 (blue) and \texttt{carmencita} (orange). The encounter with UPM J0812-3529 is spurious and was eliminated (see text).}
    \label{fig:Sun_100_BJ}
\end{figure}

First, we calculated the number of close stellar encounters with the Sun using the 100\,pc catalogue and applying the LMA described in Sect.\,\ref{sec:methodology} (see Fig.~\ref{fig:Sun_100_BJ}).

\begin{figure}
        \includegraphics[width=\columnwidth]{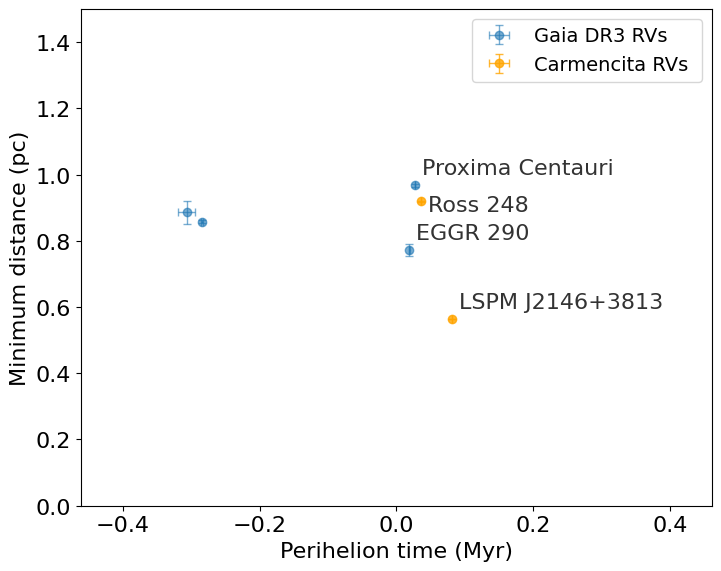}
    \caption{Same as Fig. \ref{fig:Sun_100_BJ} but using a 25\,pc limit.}
    \label{fig:Sun_25}
\end{figure}

In Fig.~\ref{fig:Sun_25}, we show the results obtained when applying the $l_{\mathrm{lim}}$ distance limit of 25\,pc. Under this assumption, the Sun experiences a total of 6 close encounters within the adopted time limits, as described in Sect.\,\ref{sec:methodology}. We excluded the encounter with UPM J0812-3529, as it does not represent a genuine close approach. This object has been classified as a white dwarf \citep{Finch2018}, presenting a strong magnetic field \citep{Bagnulo2020}. UPM J0812-3529 has a very low ratio of transverse velocity to RV in \textit{Gaia} DR3, which casts doubt on the validity of the RV. White dwarf RVs are notoriously difficult to measure due to the absence of well-defined, narrow spectral lines in their spectra \citep{Rogers2023}. In the \textit{Gaia} DR3 catalogue, the RV value of UPM J0812-3529 is $-373.7\pm8.2$\,km\,s$^{-1}$. However, \citet{Landstreet2023} concluded that this value was not correct and estimated the real RV to be $+83\pm140$\,km\,s$^{-1}$, which would indicate that the white dwarf is in fact moving away from the Sun. Since a reliable RV is not available, we excluded this target from the analysis. Six close encounters remained within the $\pm 0.47$ Myr time window and the $25$\,pc distance limit used for the statistical comparison.

\renewcommand{\arraystretch}{1.4}
\begin{table*}
\centering
\caption{Predicted stellar encounters of the Sun with systems within 25\,pc.}
\label{tab:sun_ce25}
\begin{tabular}{lclll}
\toprule
Common name &  Periastron distance (pc) & Periastron time (kyr) & Periastron velocity (km/s) \\
\midrule
\textbf{GJ~710\textsuperscript{*}} &  0.0621 $\pm$ 0.0023 & 1344.6 $\pm$ 2.2 & $-$13.899 $\pm$ 0.022 \\ \midrule
LSPM J2146+3813   & 0.56423 $\pm$ 0.00077 &     81.89 $\pm$ 0.11   &  $-$83.87 $\pm$ 0.11 \\
EGGR 290          & 0.772 $\pm$ 0.019     &     19.03 $\pm$ 0.48   & $-$415 $\pm$ 10 \\
UPM J1121-5549    & 0.8576 $\pm$ 0.0047   & $-$283.8 $\pm$ 1.5     &     76.52 $\pm$ 0.41 \\
UCAC4 213-008644  & 0.886 $\pm$ 0.036     & $-$306 $\pm$ 13        &     79.2 $\pm$ 3.2 \\
Ross 248          & 0.9187 $\pm$ 0.0012   &     35.962 $\pm$ 0.042 &  $-$82.20 $\pm$ 0.11 \\
Proxima Centauri  & 0.9679 $\pm$ 0.0045   &     26.574 $\pm$ 0.029 &  $-$32.04 $\pm$ 0.22 \\
\bottomrule
\end{tabular}
\tablefoot{$^{\star}$ GJ~710  falls outside the selected time window $t_{\mathrm{lim}}=0.47$\,Myr and is shown separately for reference.}
\end{table*}
\renewcommand{\arraystretch}{1.0}

Table ~\ref{tab:sun_ce25} lists the periastron distance, periastron time, and periastron velocity for all encounters with the Sun within the adopted time and distance limits. We note that periastron velocity corresponds to the initial velocity for all cases except GJ~710, where we integrated the trajectory instead of using the LMA method. The encounter with GJ~710 is technically outside of the $\pm 0.47$\,Myr, but we included it in the table because of the particular interest it holds and the specific discussion below. Our updated calculations predict a passage at a periastron distance of $d_{\mathrm{p}} = 0.0621\pm0.0023$\,pc from the Sun in $1344.6\pm2.2$\,kyr. For comparison, \cite{Bailer-Jones2022} reported $d_{\mathrm{p}}=0.064\pm0.0024$\,pc (median $0.0636$\,pc, 90\% confidence $0.0595 - 0.0678$\,pc) in $1292 \pm 23$\,kyr. Their analysis also integrated the trajectories through a Miyamoto-Nagai potential using \textit{Gaia} DR3 data. Our minimum distance agrees with their results, but the encounter time has been offset due to our RV corrections, reducing the initial relative velocity and delaying periastron passage time. A test using uncorrected RVs recovers a periastron time consistent with \citeauthor{Bailer-Jones2022}.

Another remarkable close encounter is that of Proxima Centauri. The predicted minimum distance of $d_{\mathrm{p}} = 0.9679\pm0.0044$\,pc will occur in $26.574\pm0.028$\,kyr. It is important to note that Proxima Centauri is not an isolated object, but a member of the $\alpha$ Cen triple system, together with $\alpha$ Cen A and B. These two components are bright solar-type stars and are not included in \textit{Gaia} DR3 due to saturation effects. Consequently, we can assume that the encounter calculated here corresponds to the entire $\alpha$ Cen system.

It is worth noting that EGGR 290 exhibits an exceptionally high velocity of $415\pm10$\,km s$^{-1}$, making it a true high-velocity `bullet' star. Its inclusion as an encounter in our calculations is only due to its proximity (within 25\,pc from the Sun). However, this type of fast-moving object would be missed if it were located beyond the 25\,pc limit, since at such speed a distance of 25\,pc is covered in just $\sim 60$\,Kyr.

\subsection{Close encounter statistics} \label{sec:Res_CE_statstics}

As described in Sect.\,\ref{sec:methodology}, we calculated encounters among all stars within our dataset. A total of 3\,765 systems are within the distance limit of $l_{\mathrm{lim}}=25$\,pc and were selected as central stars for the encounter analysis. Overall, we identified 16\,534 close encounters ($d_{\mathrm{p}}<1$\,pc), providing a robust sample for statistical analysis. We decided to investigate more closely those cases where the encounters initial distances are nearly equal to their periastron distances. The reason for that is to assess whether those correspond to resolved binary systems or common proper motion pairs (CPMPs), which are naturally flagged as encounters in our calculations.
Furthermore, additional encounters involving resolved binary systems or CPMPs can affect the global statistics depending on whether they are treated as single systems or as two separate stars. Beyond the statistical impact, encounters with multiple stellar systems also have specific physical relevance as they can lead to more dynamically complex perturbations than those involving single stars \citep{Li2015, Heggie&Rasio1996}, which may significantly alter the properties of the stellar systems.

\begin{figure}
        \includegraphics[width=\columnwidth]{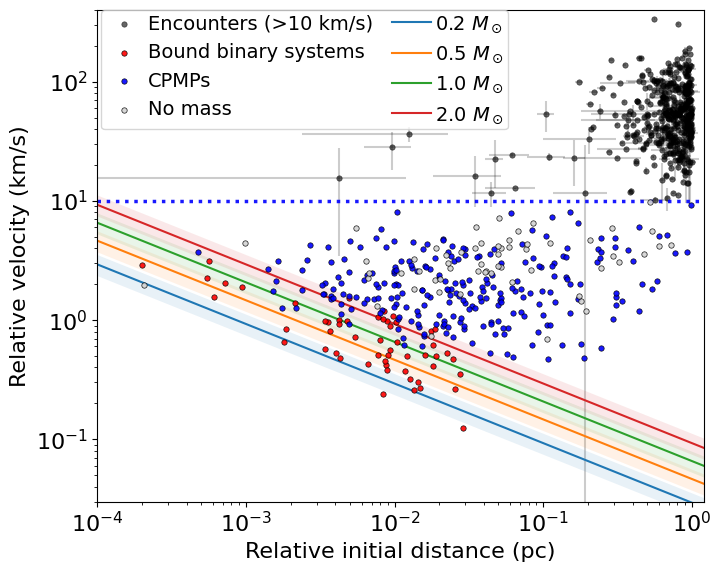}
    \caption{Relative velocity vs initial relative distance ($<$1\,pc) for the close encounters. Theoretical curves for binary systems with different total masses (assuming a circular orbit) are also displayed. The adopted 10\,km\,s$^{-1}$ threshold is shown with a dashed blue line.}
    \label{fig:Bin/CMP}
\end{figure}

Figure~\ref{fig:Bin/CMP} displays the encounter velocity (periastron speed) as a function of the initial relative distance for encounters up to 1\,pc. Two distinct regions are evident in the plot. The top-right region consists of encounters with initial distances near 1\,pc and periastron velocities of approximately 50\,km\,s$^{-1}$. We consider them to be genuine star-to-star close encounters as their periastron speeds are too large for them to be gravitationally bound. However, the broader, lower-velocity region is likely to be composed of binary systems and CPMPs. The overlaid theoretical curves represent gravitationally bound binary systems (assuming a circular orbit) where the periastron speed is treated as a constant orbital velocity and the distance corresponds to the orbital semi-major axis. Relationships for binary systems with equal-mass components are shown in different colours for various total stellar masses, as labelled. The overlap of the lower boundary of this region with the theoretical curves supports the interpretation that these slower encounters correspond to gravitationally bound binary systems or nearly bound CPMPs.

To identify potential binaries and CPMPs, we established a velocity lower limit of $v_{\mathrm{p}} \leq 10$\,km\,s$^{-1}$ for encounters occurring at initial distances $<1$\,pc. This threshold was selected conservatively after systematically testing multiple values (e.g. 5, 10, 15 km s$^{-1}$), which resulted in no significant changes to the overall encounter statistics or distributions. As shown in Fig.~\ref{fig:Bin/CMP}, the adopted strategy effectively separates two distinct regions, corresponding to systems that fortuitously have a close encounter at the present time and those that are bound or nearly bound.

As a next step, we carriedo out a classification among resolved binaries (bound) and CPMPs (unbound) by computing the ratio of the kinetic ($E_{k}$) to the potential ($E_{p}$) energy of each pair of stars, calculated as
\begin{equation}\label{eq:Eratio}
    R_{i,j} = \frac{E_k}{|E_p|} = \frac{r_{i,j} v_{i,j}^2}{2G(m_i+m_j)},
\end{equation}
where $r_{i,j}$ and $v_{i,j}$ are the relative distance and velocity between the stars, respectively, and $m_{i,j}$ are their corresponding masses. Whenever possible, we estimated the mass of the components from the available stellar radius and surface gravity ($log\ g$) measurements. For stars without these parameters, but with a known spectral type, we adopted the median mass corresponding to stars of the same spectral type.
We classified bound systems satisfying $R\leq1$ as resolved binaries and pairs with unbound energy ratio $R>1$ as CPMPs. This approach resulted in 54 binary systems, 239 CPMPs, and 58 pairs for which the mass of at least one component could not be determined.
For the binary systems, we treated the system as one single object for all subsequent analyses. To do so, we substituted the two individual components by the pair's barycentric positions and velocities. The 54 resulting single joint systems were included in the catalogue and were used to recompute the encounter calculations. For the CPMPs, we simply removed the internal encounter between the components from our statistics, but we considered further encounters as two individual objects. Systems where the mass could not be determined were treated as CPMPs.

\begin{figure*}
        \includegraphics[width=2\columnwidth]{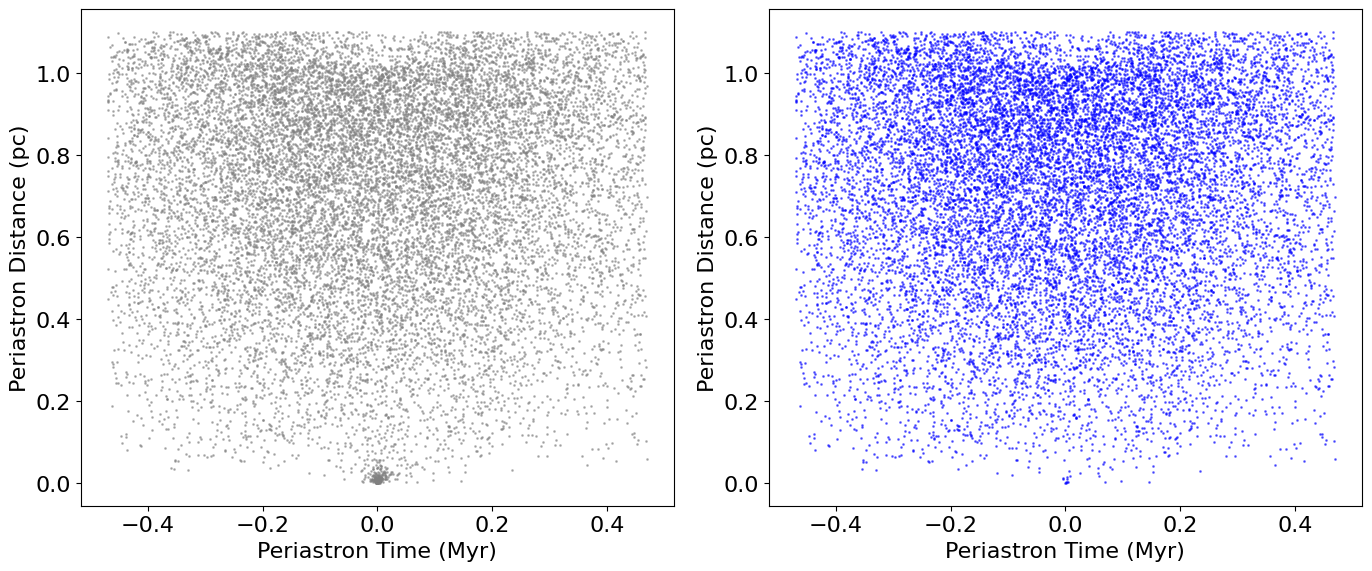}
    \caption{Periastron distance as a function of periastron time for the close encounters before (left) and after (right) the removal of binary systems and CPMPs.}
    \label{fig:dist_time-Before/After}
\end{figure*}

Figure~\ref{fig:dist_time-Before/After} illustrates the periastron distance as a function of periastron time, both before and after the removal of binary systems and CPMPs. This figure illustrates that the apparent excess of encounters occurring at very small distances (below 0.1\,pc) and with periastron times near the present time is largely corrected when applying the $v_{\mathrm{p}} \leq 10$\,km\,s$^{-1}$ threshold criterion. Our final sample consist of 15\,733 close encounters that are suitable for statistical analysis.

\begin{figure}
        \includegraphics[width=\columnwidth]{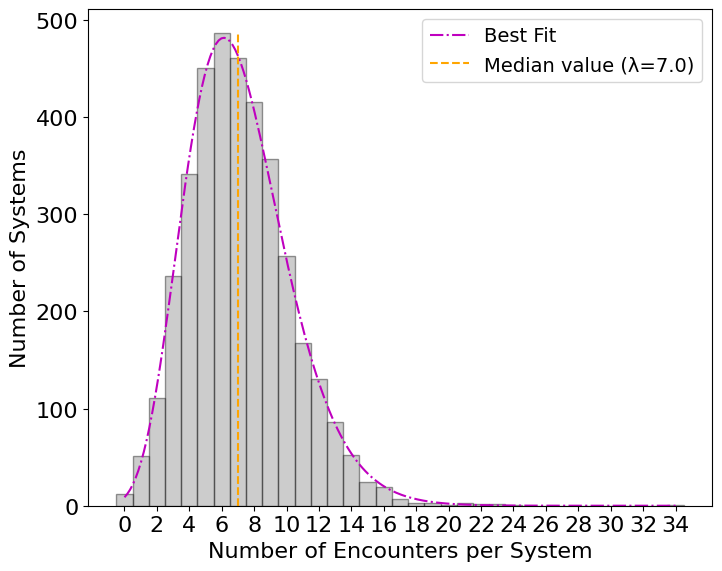}
    \caption{Histogram of the number of encounters per stellar system with the best-fitting Beta function (purple) and the median value (orange).}
    \label{fig:Number_statistic}
\end{figure}

Figure~\ref{fig:Number_statistic} shows the histogram of the number of stars experiencing a given number of close encounters within the adopted time limits. From the distribution, we find that the median number of close encounters within the considered $0.94$ Myr total time interval is $7.0 \pm 3.0$. Also note that the distribution is not symmetric. This comes from the fact that the periastron distance is a positive-defined variable. For comparison, we have calculated that the Sun experiences six close encounters within the same time and distance limits, a value that places the Sun at the mode of the distribution, indicating that its encounter frequency is typical for stars in the local neighbourhood (within 25\,pc).
As a further check of the impact of close binaries and CPMPs, we estimated the same statistics by considering all systems showing a relative velocity below 10\,km\,s$^1$ (blue dotted line in Fig.\,\ref{fig:Bin/CMP}) as bound binary stars. The results show a slightly lower median number of encounters per star ($6.0 ^{+3.0}_{-2.0}$) but the difference remains well within the uncertainties, indicating that this has a minor impact on the overall results.

\begin{figure*}
    \centering
    \begin{subfigure}[t]{0.32\textwidth}
        \includegraphics[width=\textwidth]{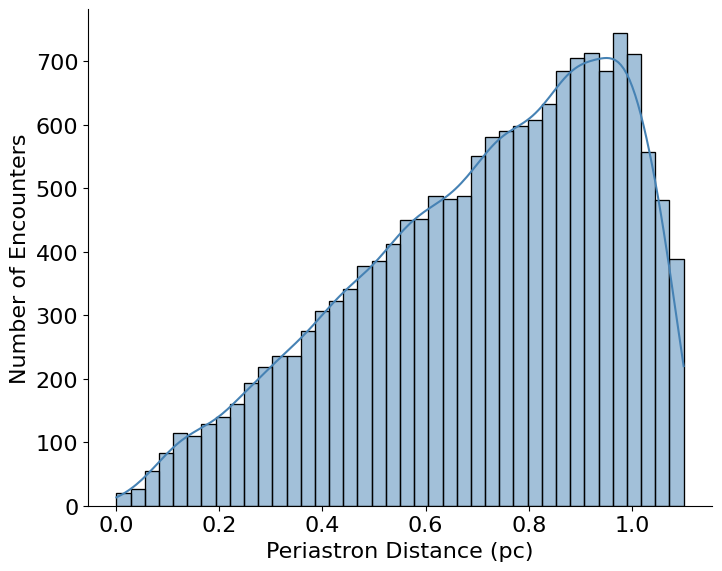}
        \caption{}
        \label{fig:dist_panel}
    \end{subfigure}
    \hspace{0.01\textwidth}
    \begin{subfigure}[t]{0.32\textwidth}
        \includegraphics[width=\textwidth]{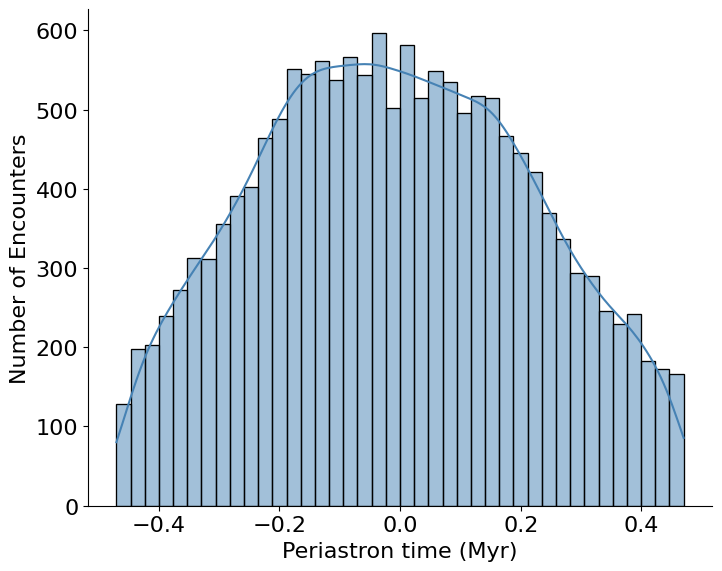}
        \caption{}
        \label{fig:time_panel}
    \end{subfigure}
    \hspace{0.01\textwidth}
    \begin{subfigure}[t]{0.32\textwidth}
        \includegraphics[width=\textwidth]{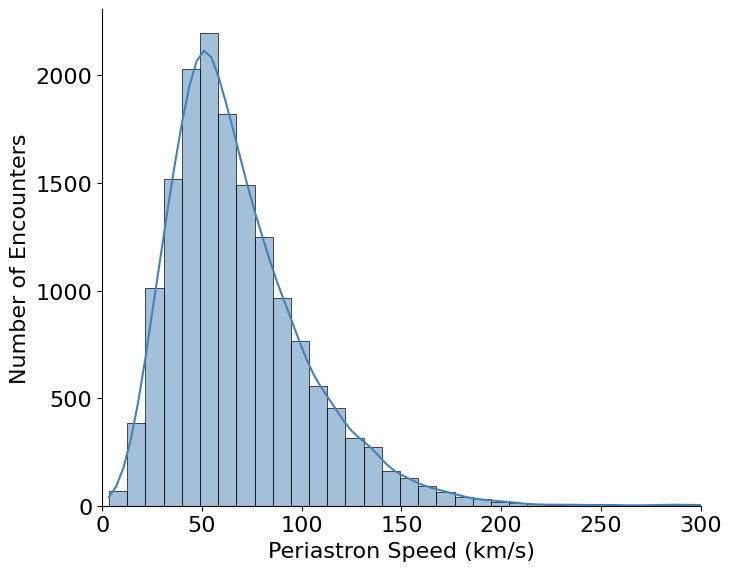}
        \caption{}
        \label{fig:speed_panel}
    \end{subfigure}
    \caption{Distributions of close encounters as a function of ($a$) periastron distance, ($b$) periastron time, and ($c$) periastron speed.}
    \label{fig:distribution}
\end{figure*}

Panel $a$ in Fig.\,\ref{fig:distribution} depicts the histogram of the number of close encounters as a function of periastron distance. As expected, the number of encounters increases with periastron distance since geometrically and statistically, close approaches at very small distances are significantly rarer than those at larger separations. Beyond the 1-pc threshold, however, the number of encounters decreases due to the construction of our sample of close encounters, which only considers cases where the periastron distance goes below 1\,pc at the 1$\sigma$ lower limit.

Panel $b$ in Fig.\,\ref{fig:distribution} shows the distribution of close encounters as a function of periastron time. Assuming a homogeneous distribution of stars in the solar neighbourhood, a uniform distribution would be expected, indicating that encounters are equally probable at any point in time. We attribute the deviation from a flat distribution to two effects: (i) increasing uncertainty in the calculated periastron distance when looking further into the past or future and (ii) contributions from fast-moving stars beyond the 25-pc limit. At times near the present ($t_{\mathrm{p}} \sim 0$), the positions and velocities of stars are well constrained, allowing for a reliable identification of all close encounters. For wider temporal windows, however, the uncertainty in periastron distance increases. While encounters within 1$\sigma$ of the 1-pc threshold are included, those with lower probabilities are progressively excluded. The second effect includes systems with very large relative velocities, such as EGGR 290, which exceeds the 90th percentile of the RV distribution (the criterion used to define our time limit; see Sect.\,\ref{sec:methodology}). In fact, Fig.\,\ref{fig:distribution} shows that we are only $\sim$100\% complete up to $\pm0.2$\,Myr, meaning that any star with a velocity exceeding $\sim122$\,km s$^{-1}$ and outside the 25\,pc sphere, will be missing from our encounter statistics. Since the number of missed close encounters near $t_{\mathrm{p}} \sim 0$ should be close to zero, we can extrapolate the encounter rate at the present time to the full time interval. 

By integrating the distribution shown in Fig.~\ref{fig:time_panel}, we find that our calculations currently capture 70\% of the encounters. This implies that close encounter rate reported in Fig.~\ref{fig:Number_statistic} underestimates the true value, which we correct to $10.6 \pm 4.5$ per Myr and star, or equivalently, one encounter every $95^{+71}_{-28}$ kyr. An independent verification of this correction can be obtained by comparing the number of encounters when adopting different distance limits for the Sun, that is, 100\,pc and 25\,pc (Fig.~\ref{fig:Sun_100_BJ} and Fig.~\ref{fig:Sun_25}. Indeed, if we consider a time interval $\pm0.47$\,Myr, the a 25-pc criterion captures approximately 70\% of the encounters found with the 100\,pc limit.

It is easy to show that the encounter rate scales quadratically with periastron distance (impact parameter). If we treat one system as a stationary target and the other as moving impactor, the effective cross-section area for the encounter scales with the square of the minimum distance considered. Therefore, the expected number of encounters also varies quadratically with the criterion adopted for the minimum encounter distance. This scaling enables an estimation of the encounter rate for a minimum approach distance of 0.5\,pc (i.e. the conventionally accepted outer limit of the Oort cloud) and we find it to be $2.6 \pm 1.1$ per Myr. During the Solar System's 4.56 Gyr lifespan \citep{Amelin2002}, the calculations predict $12\,000 \pm 5000$ stellar fly-bys within the Oort cloud. This can lead to significant dynamical perturbations, which imply object transfer between the two stars and the ejection of cometary bodies into the inner Solar System and out to interstellar space \citep{Torres2019}.

\subsection{Comparison with previous studies}

We can compare the results shown in Fig.~\ref{fig:Sun_100_BJ} with those presented by \citet{Bailer-Jones2022} who performed numerical integrations with a Galactic potential to determine the trajectories of stellar encounters with the Sun. In their work, the authors identified 61 stellar encounters within 1\,pc of the Sun across an unrestricted time window. On the other hand, our analysis yields 24 stellar encounters within the same distance but restricted to $\pm 1.85$\,Myr. Their larger number reflects the difference in the selection criteria applied to the dataset specifically, limited to 100\,pc distance from the Sun and with a parallax-over-error $>5$. Despite this, all encounters identified in our calculations are also reported by \citet{Bailer-Jones2022} and the derived periastron parameters are generally consistent between both studies. The main differences occur for three encounters in the distant future (several Myr, outside our $t_{\mathrm{lim}}$ window), where the LMA model yields results that differ significantly from the numerical integrations within a Galactic potential. 

A key improvement of our study over previous ones, is the physically motivated treatment of RV corrections, accounting for both gravitational redshift and convective blueshift. Unlike previous analyses that assumed these effects to largely cancel, we demonstrate they can yield a net positive RV shift, leading to refined and not biased periastron times, particularly for encounters far into the future or past.

The encounter rate derived from our analysis, $10.6 \pm 4.5$ per Myr and star, is in agreement with previous estimates of the encounter frequency for the Solar System. \cite{Garcia-Snchez2001} report a value of $11.7 \pm 1.3$\,Myr$^{-1}$ within 1\,pc, while \citet{Bailer-Jonesetal2018} find $19.7\pm2.2$\,Myr$^{-1}$ for encounters within 1\,pc and within 15\,Myr of the present. The lower rate we find cannot be entirely attributed to the more restrictive spatial (25\,pc) and temporal ($\pm 0.47$ Myr) limits adopted in our calculations, nor to the incompleteness caused by missing very fast encounters (faster than $v\sim122$\,km s$^{-1}$), since we apply a correction for these based on Fig. \ref{fig:time_panel}. The discrepancy is also unlikely to arise solely from the different treatment of the binary systems and CPM pairs. We therefore attribute the higher rate reported by \citet{Bailer-Jonesetal2018} mainly to the effect they acknowledge: imperfect filtering of spurious parallaxes, which can artificially increase both the number of encounters and the inferred encounter rate.

\subsection{The special case of GJ~710}

According to our updated calculations, GJ~710 will make a close approach to the Sun at only $12\,809\pm470$\,AU in $1.3446 \pm 0.0022$\,Myr. Given the statistical results of our analysis, we find that the case of GJ~710 is remarkable because of two main reasons. The encounter distance is quite extreme, as can be seen in panel $a$ of Fig.~\ref{fig:distribution}. We find that within our limited range of $\pm 0.47$\,Myr and 25\,pc, there are 57 encounters between stars that result in minimum distances below $0.0621$\,pc. Following the discussion above, this value needs to be corrected considering the fact that we are progressively missing close encounters as we approach the edges of the time interval. As discussed above, we estimate that the corrected number of encounters should rather be $81.0 \pm 7.1$. To compute the encounter rate, we consider the average number of stars within a 25\,pc sphere, which is $3656.36$. Considering the total time span of $0.94$\,Myr, we obtain an  encounter rate for approaches as close as GJ~710's of $0.0210 \pm 0.0018$ per Myr. This frequency corresponds to an average of one such extremely close encounter per star every $47.7^{+4.6}_{-3.9}$\,Myr. This shows that the GJ~710–Sun encounter happening in only 1.3\,Myr is statistically uncommon and represents an exceptional event.

The relatively low periastron velocity of GJ~710 of $-13.899\pm0.022$\,km\,s$^{-1}$ is also remarkable. It places GJ~710 at the 0.61\% percentile, very close to the lower bound of the periastron velocity distribution shown in panel \(c\) of Fig.\,\ref{fig:distribution}. This distribution shows a median velocity of $\sim61^{+38}_{-23}$\,km\,s$^{-1}$, reflecting the typical velocity dispersions of stars in the solar neighbourhood. GJ~710 will have a relatively slow fly-by, allowing for prolonged gravitational interaction and substantial dynamical perturbation on the Solar System. During this approach, GJ~710 will traverse the Oort cloud, modelled as a sphere with an outer radius of $0.50$\,pc. The chord length (the straight-line path segment through the cloud at periastron) is $L = 2 \sqrt{R^2 - d^2_p} = 0.992$\,pc, yielding a crossing time of $t_\mathrm{cross} = 69.9 \pm 0.3$\,kyr.

\section{Conclusions}
\label{sec:conclusions}

In this work, we study the statistics of close stellar encounters in the solar vicinity, paying special attention to placing the case of GJ~710’s close encounter with the Solar System in a broader context. This event has the potential to gravitationally perturb the outer regions of the Solar System, such as the Oort cloud, or even directly affect planetary orbits.

We used a sample of stars within 100\,pc from \textit{Gaia}, complemented with RVs from CARMENES, to construct a complete dataset suitable for a statistical study of close encounters among them. We carefully assessed the completeness of our dataset and applied corrections to the RVs to account for gravitational redshift and convective blueshift, which bias the measured velocities and primarily affect the derived encounter times. These improvements enabled more precise and robust encounter statistics compared to previous studies.

We have identified a total of six close encounters ($<1$\,pc) with the Sun of stars closer than 25\,pc within a time window of $\sim$1\,Myr. Expanding the analysis to study fly-bys among all stars, we conclude that the rate of close encounters in the solar vicinity is about $10.6 \pm 4.5$ per Myr and star, after correcting for incompleteness and accounting for close approaches of multiple systems. This value corresponds to one encounter every $95^{+71}_{-28}$ kyr.

We also computed the rate of encounters within 0.5\,pc (Oort cloud limit) to be $2.6 \pm 1.1$ encounters per Myr, meaning that the Solar System has experienced $\sim12\,000 \pm 5000$ such fly-bys over its 4.56\,Gyr lifetime. These perturbations drive cometary exchange and ejections into the inner Solar System and interstellar space~\citep{Torres2019}.

Finally, we examined the statistical frequency of the GJ~710–Sun encounter. Based on the resulting statistics, we find that an approach closer than the expected GJ~710 fly-by is quite rare, with an average frequency of one such event per star every approximately $50$\,Myr. However, from another perspective, stars in the solar neighbourhood come as close as $\sim 10^4$\,AU, approximately 20 times within a 1-Gyr time interval. The cumulative effect of the dynamical perturbations is potentially large. 

From an entirely different perspective, we might go on to speculate that such close encounters could be potentially used as an energy-efficient means for a civilisation to extend itself across the Galaxy. This could be dubbed a `star-hopping' strategy. With this approach, a civilisation could wait for the close passage of a star and settle on a hypothetical planet (or moon) to continue its galactic journey. This could allow for a slow, but exponential spread of a civilisation across the Galaxy by utilising minimal energy resources \citep{Carroll-Nellenback2019, Hansen2021}.

In conclusion, our study provides new insights into the broader dynamics of stellar encounters in the solar neighbourhood. The statistical relevance of the upcoming GJ~710 fly-by highlights the astrophysical importance of continued monitoring and analysis of this system, particularly with respect to the search for potential planetary companions.

\begin{acknowledgements}
Eloi Fernandez-Puig carried out this work within the framework of the doctoral program in Physics of the Universitat Autònoma de Barcelona.
We acknowledge financial support from the Agencia Estatal de Investigaci\'on (AEI/10.13039/501100011033) of the Ministerio de Ciencia e Innovaci\'on and the ERDF ``A way of making Europe'' through projects PID2021-125627OB-C31 and PID2024-158486OB-C31, from the programme ``Unidad de Excelencia Mar\'ia de Maeztu'' CEX2020-001058-M financed by MCIN/AEI/10.13039/501100011033 and the MaX-CSIC Excellence Award MaX4-SOMMA-ICE. This work was also funded by the European Research Council (ERC) under the European Union’s Horizon Europe  programme (ERC Advanced Grant SPOTLESS; no. 101140786), the Secretaria d'Universitats i Recerca del Departament d'Empresa i Coneixement de la Generalitat de Catalunya and the Ag\`encia de Gesti\'o d’Ajuts Universitaris i de Recerca of the Generalitat de Catalunya, with additional funding from the European FEDER/ERF funds, \emph{L'FSE inverteix en el teu futur}, from the Generalitat de Catalunya/CERCA programme. This work has made use of data from the European Space Agency (ESA) mission \textit{Gaia} (\url{https://www.cosmos.esa.int/gaia}), processed by the \textit{Gaia} Data Processing and Analysis Consortium (DPAC, \url{https://www.cosmos.esa.int/web/gaia/dpac/consortium}). Funding for the DPAC has been provided by national institutions, in particular the institutions participating in the \textit{Gaia} Multilateral Agreement.

\end{acknowledgements}


%

\bibliographystyle{aa} 
\bibliography{aa59497-26} 

\begin{appendix}




%

\clearpage

\end{appendix}
\end{document}